\documentclass{appolb}
\usepackage{graphicx}
\usepackage{mathtools}

\newcommand{\CA}{{\cal A}}

\newcommand{\CD}{{\cal D}}

\newcommand{\CI}{{\cal I}}
\newcommand{\CJ}{\cal J}

\newcommand{\CR}{{\cal R}}

\newcommand{\CQ}{{\cal Q}}

\newcommand{\average}[1]{\left\langle #1 \right\rangle_\CD}

\newcommand{\caverage}[1]{\left\langle #1 \right\rangle_{\CI}}

\newcommand{\initial}[1]{{{#1}_{\mathbf i}}}

\newcommand{\inI}{{\mathrm{I}}}
\newcommand{\inII}{{\mathrm{II}}}
\newcommand{\inIII}{{\mathrm{III}}}

\begin{document}
\title{A few numbers from the turnaround epoch of collapse%
}
\author{Jan J. Ostrowski 
\address{Department of Fundamental Research,
  National Centre for Nuclear Research, Pasteura 7, 02-093 Warszawa, Poland}
\\
}

\maketitle
\begin{abstract}
The turnaround epoch of gravitational collapse is examined by means of relativistic Lagrangian perturbation theory. Averaged, scalar equations applied to the fluid's evolution reveal some scale-independent universality of parameters for a wide variety of initial conditions. In particular, the density contrast of the collapsing domain at the turnaround is shown to be significantly smaller than the value provided by Eulerian perturbative (homogeneous and spherical) model. Combined curvature and kinematical backreaction are shown to be of the order of the energy density. Possible improvements of our treatment are put into perspective.     
\end{abstract}
  
\section{Introduction}
Attempts to realistically model the gravitational collapse of cosmological fluids fall under few categories, the two most popular being: exact solutions to Einstein equations and perturbative approaches. Here we utilize both, in a sense that we use the specific perturbation scheme to close the averaged Einstein equations. Application of exact solutions to structure formation is often limited, either by underlying space-time symmetries or by the size of the set of admissible initial conditions. In other words, exact solutions usually lack generality. One way out of this situation is to use the averaged version of the Einstein equations (trace part) and thus reduce the complicated distribution of the energy density and geometry in some arbitrary domain to its Riemannian volume integrals. Such averaging operation, however does not commute with the time evolution. In the following, we will use the Buchert equations which take into account this non-commutativity by introducing new terms to the scalar evolution equations. As a closure condition for these equations we will use a restricted version of the first order general-relativistic Lagrangian perturbation theory solution which is basically a prescription of the fluid's local volume deformation with respect to an assumed background-volume evolution (see \cite{RZA1}, \cite{RZA2} and references therein for detailed explanation of this formalism). In our examination of gravitational collapse we will restrict ourselves to the turnaround epoch i.e. a moment when initially expanding domain starts collapsing and departing from Hubble expansion, extending the derivation in \cite{flat}.
\section{Averaged equations and Relativistic Zel'dovich Approximation (RZA)}
Starting from this section and throughout the article we will consider irrotational dust as a matter model and a synchronous, comoving foliation frame. 
The expansion tensor in this foliation reads: $\Theta^i_{\;\;j} = \frac{1}{2}g^{ik}\dot{g}_{kj}$, where $g_{ij}$ is the spatial metric and the overdot denotes derivative wrt proper time of the fluid. 
The scalar invariants of the expansion tensor read:
\begin{equation}
  \label{eq:invariants}
\inI = \mathrm{tr}\;\left(\Theta_{ij}\right)\;\;,\;\;\inII = \frac{1}{2}\left(\left(\mathrm{tr}\Theta_{ij}\right)^2 - \mathrm{tr}\left(\left(\Theta_{ij}\right)^2\right)\right)\;\;,\;\;\inIII = \det\left(\Theta_{ij}\right) \;.
\end{equation}
Domain-dependent (we will represent this feature with $\CD$ subscript) evolution of the dust is described by the Buchert equations:
\begin{equation}
  \label{buchert}
3\frac{\ddot{a}_{\CD}}{a_{\CD}}+4\pi G \average{\rho} = \CQ_{\CD}\;\;,\;\;H^2_{\CD} -\frac{8\pi G}{3}\average{\rho} = -\frac{\average{\CR}+\CQ_{\CD}}{6}\;,
\end{equation}
where angular brackets stand for integral over Riemannian volume element divided by total volume, $a_{\CD}$ is the effective scale factor equal to the cubic root of the time-dependent volume-to-initial volume ratio, $H_{\CD}$ is the Hubble parameter, $\rho$ is the energy density and $\CR$ is the 3-Ricci scalar curvature.
Term $\CQ_{\CD}$ is called the kinematical backreaction and can be expressed in terms of expansion tensor invariants:
\begin{equation}
\CQ_{\CD} = 2\average{\inII}-\frac{2}{3}\average{\inI}^2\;.
\end{equation}
The second equation of eq. (\ref{buchert}), i.e. the averaged Hamiltonian constraint can be recast into more familiar form:
\begin{equation}
\Omega^{\CD}_m+\Omega^{\CD}_{\CQ}+\Omega^{\CD}_{\CR}=1 \;,
\end{equation}
where the dimensionless parameters are defined in the following way:
\begin{equation}
\Omega_m^{\CD}=\frac{8\pi G}{3H_{\CD}^2}\average{\rho}\;\;,\;\;\Omega_{\CQ}^{\CD}=-\frac{\CQ_{\CD}}{3H_{\CD}^2}\;\;,\;\;\Omega_{\CR}^{\CD}=-\frac{\CR_{\CD}}{3H_{\CD}^2}\;.
\end{equation}
Closure of these equations can be obtained by assuming that fluid particles follow the perturbed (wrt to the fixed background) worldlines. The basic quantity we will need is the local, peculiar volume deformation:  
\begin{equation}
{\CJ} = 1+\xi(t)\inI_{{\bf i}}+\xi^2(t)\initial{\inII}+\xi^3 \initial{\inIII}\;,
\end{equation}
built from the averaged, initial, peculiar-volume invariants (with background subtracted), and $\xi(t)$ which is the background dependent growth function.
By combining the Raychaudhuri equation and the Hamiltonian constraint we obtain the expression for the averaged spatial curvature scalar:
\begin{equation}
  \label{curv}
\average{\CR} = -\left(\frac{\caverage{\ddot{\CJ}}}{\caverage{\CJ}}+ 3 \left(\frac{\ddot{\xi}}{\dot{\xi}}+4\frac{\dot{a}}{a}\right)\frac{\caverage{\dot{\CJ}}}{\caverage{\CJ}}\right)\;.
\end{equation}
In addition, first two invariants of peculiar-expansion tensor read:
\begin{eqnarray}
  \label{invrza}
  \average{\inI(\theta^i_{\;j})}&=&\frac{\caverage{\dot{\CJ}}}{\caverage{\CJ}}\;\;,\;\;\average{\inII(\theta^i_{\;j})}=\frac{1}{2}\left(\frac{\caverage{\ddot{\CJ}}}{\caverage{\CJ}}-\frac{\ddot{\xi}(t)}{\dot{\xi}(t)}\frac{\caverage{\dot{\CJ}}}{\caverage{\CJ}}\right)\;\;,
  \end{eqnarray}
where $a$ is the background scale factor and we used another averaging operator:
\begin{equation}
\average{\CA}=\frac{\caverage{\CA \CJ}}{\caverage{\CJ}}
\end{equation}
\section{Results: turnaround epoch of the gravitational collapse}
Cosmological compact objects (e.g. galaxy clusters), if initially expanding, must have undergone a turnaround period i.e. a period of  change in sign of the expansion scalar. The turnaround condition reads: $H_{\CD} = 0$. For simplicity, we will assume the Einstein-de Sitter background (EdS). In this case $\xi(t) = (a(t)-a(t_i))/\dot{a}(t_i)$. Using the expression for the second invariant (eq. (\ref{invrza})) we get from eq. (\ref{curv}):
\begin{equation}
\average{\CR} = -2\average{\inII(\theta^i_{\;\;j})}-\frac{\caverage{\dot{\CJ}}}{\caverage{\CJ}}\left(4\frac{\ddot{\xi}}{\dot{\xi}}+12\frac{\dot{a}}{a}\right)\;.
\end{equation}
Second term for EdS background reads (using Friedmann equations):
\begin{equation}
4\frac{\ddot{a}}{\dot{a}}+12\frac{\dot{a}}{a} = 10 H\;,
\end{equation}
where $H$ is the background Hubble parameter.
Invoking the turnaround condition:
\begin{equation}
H_{\CD}=0 \to \frac{\caverage{\dot{\CJ}}}{\caverage{\CJ}} = -3H\;,
\end{equation}
we finally obtain:
\begin{equation}
\average{\CR} = -\left(2\average{\inII(\theta^i_{\;\;j})}-30H^2\right)\;.
\end{equation}

From the averaged Hamiltonian constraint and using $H_{\CD} = 0$ we get:
\begin{equation}
\average{\CR}+\CQ_{\CD} = 24H^2 = 16\pi G \average{\rho}
\end{equation}
and thus, the density contrast and the combined backreaction and curvature--density ratio at the turnaround is ($b$ denotes the EdS normalization):
\begin{equation}
\frac{\average{\rho}}{\rho_H} = 4 \;\;,\;\;\left(\Omega^{\CD;b}_{\CQ}+\Omega^{\CD;b}_{\CR}\right)/\Omega^{\CD;b}_m=-1 \;,
\end{equation}


\section{Conclusions}
In this work we extended the analysis in \cite{flat} by relaxing the $\average{\inII_{{\bf i}}}=\average{\inIII_{{\bf i}}}=0$ condition. Our results show that the density threshold needed for the collapsing domain to reach the turnaround is lower ($\average{\rho}/\rho_H=4)$ than the one obtained from Eulerian perturbation theory ($\average{\rho}/\rho_H\approx 5.55$) and is scale-independent. Combined averaged curvature and kinematical backreaction contribute to the energy budget at the level of matter and therefore should not be neglected. Our determination of the curvature is based on the specific formula and is not unique. Comparison of the RZA formulas for curvature with exact solutions is a subject of ongoing work.  
\section{Acknowledgements}
This work was partially supported by the European Research Council (ERC) under the European Union’s Horizon 2020 research and innovationprogramme (grant agreement ERC advanced grant 740021–ARTHUS, PI: T. Buchert). Author wish to thank T. Buchert and B. F. Roukema for useful comments.

\end{document}